\documentclass[reprint,longbibliography,
 amsmath,amssymb,
 prd,superscriptaddress]{revtex4-1}
\usepackage{graphicx}
\usepackage{dcolumn}
\usepackage{bm}
\usepackage[colorlinks=true, citecolor=blue, linkcolor=blue, urlcolor=blue]{hyperref}
\usepackage{xcolor}
\usepackage{amsmath, amssymb, amsfonts}
\usepackage{mathtools}
\usepackage{subfigure}
\usepackage{mathrsfs}
\usepackage{sidecap}
\usepackage{verbatim}

\definecolor{purple1}{rgb}{128,0,128}

\usepackage{bigints}
\newcommand{\bea}{\begin{eqnarray}}
\newcommand{\ea}{\end{eqnarray}}

\definecolor{darkpastelgreen}{rgb}{0.01, 0.75, 0.24}

\def\d{\mathrm{d}}

\begin{document}

\title{On the existence of steady-state black hole analogues in finite quasi-one-dimensional Bose-Einstein condensates} 
\author{Caio C. \surname{Holanda Ribeiro}}
\affiliation{Seoul National University, Department of Physics and Astronomy, Center for Theoretical Physics, Seoul 08826, Korea} 
\author{Sang-Shin Baak}
\affiliation{Seoul National University, Department of Physics and Astronomy, Center for Theoretical Physics, Seoul 08826, Korea} 
\author{Uwe R. Fischer}
\affiliation{Seoul National University, Department of Physics and Astronomy, Center for Theoretical Physics, Seoul 08826, Korea} 

\date\today

\begin{abstract}

We theoretically propose a finite-size quasi-one-dimensional Bose-Einstein condensate with coherent source and drain placed at its two ends,  
which can in principle sustain a stationary sonic black hole with a single event horizon. 
Our analysis is focused on the condensate persistence against quantum fluctuations. 
We show that similar to white hole-black hole pairs, dynamical instabilities occur. Investigating in detail the instabilities' dependence on the system parameters, we also identify windows of formally infinite black hole lifetimes. By using quantum depletion of the condensate as a diagnostic tool, we validate the usage of Bogoliubov theory to describe the analogue Hawking process, and establish novel signatures of Hawking 
radiation in the depleted cloud, both inside and outside the event horizon.

\end{abstract}

\maketitle



\section{Introduction}

The discovery by Stephen Hawking that black holes, quantum mechanically, are not black but radiate a thermal spectrum of particles \cite{Hawking1974,Hawking1975} 
continues to furnish an intriguing milestone in the quest for a unification of quantum mechanics with gravity 
\cite{GiddingsRoyal}. 
While an 
observation of the Hawking effect  with astrophysical black holes is essentially impossible, analogue systems in fluids have enabled it due to its kinematical nature \cite{Unruh1981,Visser1998,MattCQG,BLV} and robustness
against (most variants of) Lorentz invariance breaking \cite{Jacobson1991,Unruh1995,Corley1,Universality}. 
In particular, Bose-Einstein condensates have been identified as 
suitable system to verify an analogue of Hawking's prediction in a superfluid of very low temperature, 
and other quantum effects related to sonic horizons see, e.g.  
\cite{Ted,Finazzi,PhysRevLett.85.4643,Fedichev2003,NastyPetya,Ralf2006,Carusotto_2008,Recati,Macher,Lahav,Steinhauer16,Gooding,Leonhardt2020}.

An unambiguous confirmation of the quantum Hawking effect 
was achieved in 2019 by the Steinhauer group \cite{Steinhauer2019}, 
see also the more recent experiment \cite{Steinhauer2021}.
The observation of the density-density correlations as a second-order correlation function signature of the Hawking effect \cite{Carusotto_2008,Balbinot2008,SteinhauerPRD}, 
however, still presents, in particular for small 
Hawking temperatures $T_{\rm H}$, a formidable task \cite{LeonhardtAnnalen,Wang2}.

Current experiments \cite{Steinhauer2019,Steinhauer2021} are carried out with 
flow geometry of strongly elongated condensates, a primary motivation being to avoid turbulence 
developing when the condensate flows supersonically. Two aspects to be reconciled are of fundamental importance when modeling such black hole analogues, namely what is actually feasible at a laboratory level and what the mathematical complexity of theoretical models requires to be solved. For definiteness, our guiding experimental parameters for system sizes and number of condensed atoms are the ones typically currently 
implemented in the experiments of \cite{Steinhauer2019,Steinhauer2021}. In particular, the condensate is radially trapped, operating near a quasi-one-dimensional (quasi-1D) regime. 
In order to describe the phenomena presented by such quasi-1D analogues, a working hypothesis commonly assumed is the negligibility of the finite axial 
size of the condensate. When considering the Hawking process, this is usually considered 
a justified hypothesis if the system boundaries are ``sufficiently distant'' from the analogue event horizon. 
Due to the inherent complexity of analogue black holes in the many-body context of interacting condensates,  
it is however in general not possible to decide whether or not theoretical models based on the assumption of an  infinitely extended quasi-1D condensate correctly capture all features of real finite-size condensates in a controlled manner.

In the following, we employ the idea of using coherent sources \cite{Garay2001} to study a finite size quasi-1D analogue model containing only a single analogue event horizon. Our model assumes the existence of a flowing condensate which is continuously pumped into the system at one of its ends and destroyed at the other, and, arguably, it represents the simplest realizable black hole analogue that captures finite size effects without the presence of a white hole. Although being experimentally more intricate than the current implementation of \cite{Steinhauer2019,Steinhauer2021}, the technology needed to sustain such a flowing 
condensate from condensate reservoirs has been established previously \cite{Santos2001,Shin,Markus,Robins2008}, the major problem being the reservoir replenishment. 
Although the latter is necessary for steady-state applications like atom lasers \cite{Robins2008,Florian}, the analogue model has to operate only for short periods of time until measurements can be performed and 
replenishment is therefore less crucial than for lasing operation.

For dilute Bose-Einstein condensates, 
an observable of fundamental relevance is quantum depletion,  
which is a 
{\em first-order} correlation function. 
We present below, to the best of our knowledge, the first calculation of depletion in inhomogeneous BECs with a sonic spacetime horizon, to certify whether under certain conditions 
 single-horizon finite-size analogue black holes can be prepared in a quasi-1D condensate.
We stress in this regard that ost of the current models for 1D analogue black holes implement infinitely extended quasicondensates that 
break the validity of Bogoliubov theory by leading to a depletion diverging everywhere
with axial system size, a well known fact for any 1D system to which the 
$f$-sum rule can be applied \cite{Hohenberg}. 
In our 1D model, depletion is everywhere finite and small for typical black hole parameters throughout the system evolution.
We thereby validate the Bogoliubov expansion, which imposes as a prerequisite that depletion must be small. 

Although challenging to validate experimentally, that analogue Hawking radiation exists is unquestionable by 
very general arguments \cite{Visser1998}, and perhaps more important is what analogue gravity theory in 
quantum many-body (condensed matter) systems can teach us further than predicting the very existence of Hawking radiation per se. 
Of central importance in this regard is how quantum (and thermal)
fluctuations created by the Hawking and other processes, propagating on the top of the condensate background backreact on this background, and thus on the motion of the condensate, which in turn affects the production of Hawking radiation. This backreaction can only be described properly within a number-conserving formalism \cite{Fischer2005}.  
The determination of quantum depletion is a first necessary step in the complex backreaction program.

The model explored here shares some dynamical features with analogue black hole-white hole systems built from toroidally flowing condensates. Namely, a finite size-induced field dynamical instability develops
in such models as well. We therefore review field quantization in the presence of instabilities, which is a well understood topic explored in a plethora of physical contexts, from condensed matter systems \cite{Leonhardt2003,Coutan2010,Ribeiro2020} to cosmology \cite{Lima2010,Lima20102}. In this context, of particular importance for the Bogoliubov theory is the time-translation-symmetry spontaneous breakdown in stationary condensates by the growing vacuum fluctuations, which prevents 
the existence of a preferred instantaneous vacuum state. We address this problem by constructing solutions to the Bogoliubov-de Gennes equation which takes into account as a starting point of the black hole formation process 
a well-defined quasiparticle vacuum state.

Describing the dynamical instabilities, our major findings include the simulation of black hole lifetimes as function of experimental parameters, whose intricate functional dependence reveals how strong the correlation of 
finite size and stability of the black hole is. We also demonstrate the existence of stability regions 
in parameter space, which black and white hole pairs was obtained previously in \cite{Garay2001}.  It is demonstrated that 
when the sonic horizon just emerges, distinct depletion signatures appear. In particular, we show that the Fourier spectrum of the depletion changes and new peaks can be detected, which constitutes a valuable tool to identify the emergence of quantum Hawking channels when negligible flux at infinity is present, and the 
conventional density-density correlation signatures are too weak to be detected by using {first-order} 
correlations. Moreover, we show that the radiated signal is correlated with the depletion cloud outside the black hole, and our simulations reveal that the radiative process is accompanied by an increase of the local depletion. 
 

\section{The model}
\label{secgtilde}

Within the s-wave approximation, 
the one-dimensional Bose-Einstein condensate under study is described by the 
action functional ($\hbar=m=1$)
\begin{equation}
S_{\rm c}=\int\d^2x\Psi^{*}\left[i\partial_t+\frac{\partial_x^2}{2}-U_{\rm e}-\frac{g}{2}|\Psi|^2\right]\Psi,\label{actionc}
\end{equation}
where $U_{\rm e}$ is the external potential. Our goal is to study a black hole model which captures the finiteness of condensates while enabling a fully analytical treatment of the quantum field operator expansion, and which contains as a limiting case an infinite size (quasicondensate) black hole analogue. 
There exists different routes for building such confined flowing condensates.  
In the experiments of \cite{Steinhauer2019,Steinhauer2021}, a condensate initially at rest is subjected to a moving blue-detuned laser and an analogue event horizon is thereby created dynamically. The drawback of this model is that 
notions of a stationary regime are difficult to establish and a fully numerical analysis is therefore unavoidable. In particular, the moving horizon is responsible for the emergence of a inner horizon, and a black hole-white hole pair forms \cite{Wang1,Wang2,Steinhauer2021}. Different techniques possible to establish analogue event horizons include the condensate being released from a reservoir \cite{de_Nova_2014} by an outcoupler, and the flowing condensate in toroidal configurations \cite{Garay2001}, the latter always containing a black hole-while hole pair, as dictated by the very ring topology.

The strategy assumed here consists in building a flowing condensate of finite size sustained by continuous coherent sources and drains \cite{Shin,Garay2001} at its boundaries. Mathematically, these are generically modeled by adding to the action \eqref{actionc} a 
term
\begin{equation}
S_{\rm s}=\int\d^2x(J_{\rm e}\Psi^*+J_{\rm e}^*\Psi),
\end{equation}
where $J_{\rm e}$ represents the external sources (and drains). Variation of the total action $S_{\rm c}+S_{\rm s}$ with respect to $\Psi^*$ leads to the inhomogeneous Gross-Pitaevskii (GP)  equation \cite{Tobias2005}
\begin{equation}
\left(-i\partial_t-\frac{\partial_x^2}{2}+U_{\rm e}+g|\Psi|^2\right)\Psi=J_{\rm e}.\label{gpj}
\end{equation}

Before presenting the model, we expand further why it is important for our analysis to assume such finite configurations over infinitely extended models. Bose-Einstein condensates cannot exist in infinitely extended quasi-1D black hole models. This can be read, for instance, from the condensate perturbations of \cite{Pavloff2012}, which imply a (generic) logarithmic divergence of the quantum depletion with the system size at $T=0$. At finite temperature, this divergence is stronger (linear in system size), as dictated by the Hohenberg theorem \cite{Hohenberg}, showing that finite temperature effects as predicted by quasicondensate models might not be extendable to condensates. Nonetheless, henceforth our analysis is restricted to zero point (vacuum) fluctuations ($T\sim0$), which pertain to the theory sector responsible for the quantum Hawking process. In this regime, a (weak) logarithmic divergence with the system size means that we can safely consider larger condensates while maintaining suficient control of the system depletion. 

Furthermore, another crucial aspect of infinite size models is linked to the global $U(1)$ symmetry (in the absence of external sources). As the condensate existence breaks this symmetry, the theory always admits at least one zero frequency (Goldstone) excitation, and if the condensate is infinite in size, then the system spectrum is continuous. This is particularly important for black hole analogues, as the Hawking-like process is a low energy phenomenon, and thus more sensitive to boundary conditions. Accordingly, to assume robustness 
of the Hawking process with respect to the system size is a rather strong assumption that needs ramification. Indeed, the spectrum cannot be continuous for finite size configurations, and because the system is not homogeneous, a nontrivial filtering of the excitations which would exist in infinite analogues should occur.

\begin{figure}[b]
\includegraphics[scale=0.9]{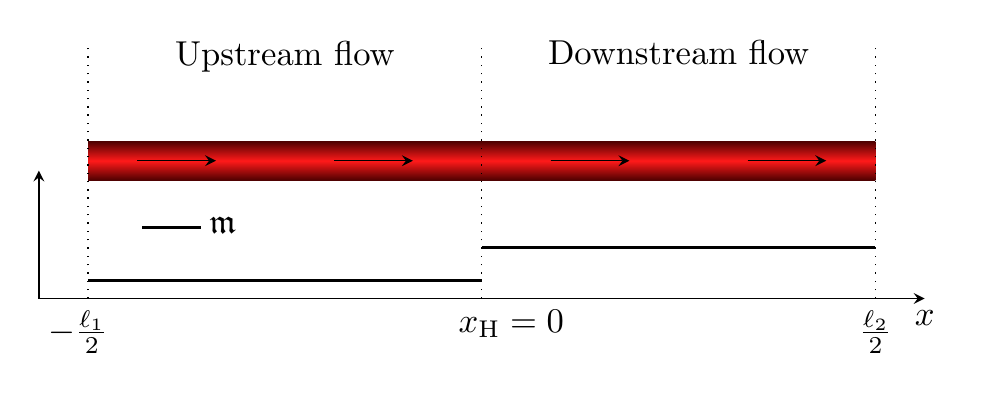}
\caption{Schematics of the condensate under study, which is assumed to be a homogeneous quasi-1D condensate of size $(\ell_2+\ell_1)/2$ flowing at constant velocity. The gas flow is sustained by continuous source 
and drain at $x=-\ell_{1}/2$ and $x=\ell_{2}/2$, respectively. At $x=x_{\rm H}$, the Mach number $\mathfrak{m}$ has a jump-like discontinuity, separating the system into two regions of different sound velocity.} 
\label{figmain1}
\end{figure}

In order to keep the analysis as straightforward as possible, we let $\rho$ be the condensate density, which is assumed to be constant in condensate support region. 
The contact interaction strength $g$ is a piecewise constant function, defined by $g=g_{\rm u}$ for $x<0$ and $g=g_{\rm d}$ for $x>0$, describing a sound barrier thoroughly studied in the literature \cite{Balbinot2008,Pavloff2012,Curtis}. The subscripts ``$\rm u$'' and ``$\rm d$'' henceforth denote ``upstream'' and ``downstream'', respectively. The condensate is trapped inside a 1D box of size $(\ell_{2}+\ell_{1})/2$ (see Fig.~\ref{figmain1}).

Now the flowing condensate is modeled by letting $\Psi=\sqrt{\rho}\exp(-i\mu t+i vx)$, where $\mu$ is the chemical potential and $v$ is the fluid velocity. By scaling $J_{\rm e}=\exp(-i\mu t+i vx)\tilde{J}/\sqrt{\rho}$, the GP equation \eqref{gpj} reduces to 
\begin{align}
&\partial_x\rho v=-2\mbox{Im}\ \tilde{J},\label{gp1}\\
&U_{\rm e}+g\rho-\mu+\frac{1}{2}v^2-\frac{\partial_x^2\sqrt{\rho}}{2\sqrt{\rho}}=\frac{1}{\rho}\mbox{Re}\ \tilde{J},\label{gp2}
\end{align}
whereas for the condensate density $\rho$ is a nonzero constant inside the box. Equation \eqref{gp1} 
represents the flux implemented by the coherent source and drain, i.e. it 
models the flux of particles fed into and removed from the condensate at $-\ell_1/2$ and 
$\ell_2/2$, respectively. Moreover, Eq.~\eqref{gp2} fixes the external potential $U_{\rm e}=\mu -g\rho-v^2/2$, necessary to sustain the condensate inside the trapping box potential. One of this model's advantages is the increased control over the system perturbations and boundary conditions (and hence over the radiation process) allowed by having available the external potential  
at the box walls (Fig.~\ref{figmain1}). We quote \cite{Vincenzo} for a comprehensive study of the various types of confining potentials that can be selected, and for our purposes we henceforth assume that $U_{\rm e}$ imposes the vanishing of the wave function at the condensate walls (corresponding to Dirichlet boundary conditions), thus fixing the non-flux part of the external source and drain through Eq.~\eqref{gp2}. Idealizing sources and potentials  as point-like simplifies considerably the analysis by enabling explicit calculations. 
The key assumption of our model is that one is able to place coherent sources and drains close to regions where the condensate is subjected to strong trapping potentials, which then prevent any condensate leakage.

The analogue model is obtained by studying sound propagation over this condensate solution, i.e., we write $\Psi=\exp(-i\mu t+iv x)(\sqrt{\rho}+\psi)$, where $\psi$ denotes freely propagating small fluctuations. This means that $\psi$ represent linearized solutions ($|\psi|^2\ll\rho$) to the homogeneous (sourceless) part of Eq.~\eqref{gpj} 
\begin{equation}
i\partial_t\psi=\left(-\frac{\partial_x^2}{2}-iv\partial_x\right)\psi+g\rho(\psi+\psi^*),\label{BdG}
\end{equation}
subjected, as mentioned, to Dirichlet boundary conditions: $\psi|_{x=-\ell_1/2}=\psi|_{x=\ell_2/2}=0$. 

A relevant quantity that can be constructed from the system parameters is $c=\sqrt{g\rho}$, which has dimensions of velocity, which  represents the local sound speed. We say that an analogue black hole condensate background
{\it exists} when $v/c_{\rm u}<1<v/c_{\rm d}$, or in terms of the Mach number, $\mathfrak{m}_{\rm u}<1<\mathfrak{m}_{\rm d}$ (Fig.~\ref{figmain1}). Thus, we need to specify $\{\mathfrak{m}_{\rm u},\mathfrak{m}_{\rm d},\ell_{1},\ell_{2}\}$, to determine the black hole completely. Also, we from now on work in units such that $c_{\rm u}=1$, which, in addition to the conventional $\hbar=m=1$, renders the upstream healing length $\xi_{\rm u}=1/\sqrt{g_{\rm u}\rho}$ to be unity. Finally, canonical quantization is obtained by promoting $\psi$ to an operator-valued distribution $\hat{\psi}$ subjected to equal-time bosonic commutation relations $[\hat{\psi}(t,x),\hat{\psi}^\dagger(t,x')]=\delta(x-x')$. The quantization details are provided in the next section.   

\section{Canonical quantization}
\label{cq}

The procedure to build the quantum field expansion for $\hat{\psi}$ follows the general recipe: solve for the quasiparticle modes, which constitute a complete set of solutions to the classical field equation; write down the most general classical solution in terms of this complete set; postulate the canonical commutation relations.  
We shall review these steps below to some degree of detail in order to fix our adopted notation.

\subsection{Quantization}

We start by defining the Nambu spinor $\Phi=(\psi,\psi^*)^{\rm t}$, where ``$\rm t$'' stands for transpose. Thus, the field equation \eqref{BdG} implies
\begin{equation}
i\sigma_3\partial_t\Phi=\left(-\frac{\partial_x^2}{2}-i\mathfrak{m}_{\rm u}\sigma_3\partial_x+\frac{g}{g_{\rm u}}\sigma_4\right)\Phi,\label{bogo}
\end{equation}
where $\sigma_i$, $i=1,2,3$ denote the usual Pauli matrices, and $\sigma_4=1+\sigma_1$. By definition, the spinor $\Phi$ satisfies $\Phi=\sigma_1\Phi^*$, and clearly, upon quantization, we should have 
\begin{align}
&[\hat{\Phi}_{a}(t,x),\hat{\Phi}^{\dagger}_{b}(t,x')]=\sigma_{3,ab}\delta(x-x'). \label{ccr}
\end{align}
Also, from the discussion after Eq.~\eqref{BdG}, we have that the field $\Phi$ is subjected to Dirichlet boundary conditions
\begin{align}
&\Phi|_{x=-\ell_{1}/2}=0=\Phi|_{x=\ell_{2}/2} .\label{bc1}
\end{align}
Furthermore, because of the boundary conditions, Eq.~\eqref{bogo} implies that if $\Phi$ and $\Phi'$ are two distinct solutions of the latter equation, then
\begin{equation}
\langle\Phi,\Phi'\rangle=\int\d x\Phi^{\dagger}(t,x)\sigma_3\Phi'(t,x)\label{scalar}
\end{equation}
is a conserved (in time) quantity, which will be used as a scalar product on the space of classical solutions. Also, as the field modes have compact support, they have finite norms, which can be taken in general as
\begin{equation}
\langle\Phi,\Phi\rangle=\pm1.\label{norm}
\end{equation}
We stress that even though Eq.~\eqref{bogo} may admit nonzero solutions with vanishing norm, we can {\it always} find an orthonormal basis as in Eq.~\eqref{norm}. The plus and minus signs in Eq.~\eqref{norm} correspond to positive and negative norm modes, and we recall that for each solution $\Phi$ of Eq.~\eqref{bogo}, $\sigma_1\Phi^*$ is also a solution of opposite norm sign.
Thus there exists a one-to-one correspondence between positive and negative norm modes, which allows us to index the positive norm solutions as $\Phi_n$, $n=1,2,3,\ldots$. With this, we can write the most general classical solution of Eq.~\eqref{bogo} as
\begin{equation}
\Phi(t,x)=\sum_{n=1}^{\infty}\left[a_{n}\Phi_{n}(t,x)+b^{*}_{n}\sigma_1\Phi_{n}^*(t,x)\right],\label{qexpansion}
\end{equation}
and in view of the reflection property $\Phi=\sigma_1\Phi^*$, it follows that $b_{n}=a_{n}$. Now, canonical quantization is defined by the promotion of $\Phi$ to the operator-valued distribution $\hat{\Phi}$ subjected to the condition \eqref{ccr}, which corresponds to promoting each $a_{n}=\langle\Phi_{n},\Phi\rangle$ to an operator $\hat{a}_{n}$ satisfying
\begin{equation}
[\hat{a}_{n},\hat{a}^{\dagger}_{n'}]=\delta_{n,n'}.\label{ccr2}
\end{equation}
Concluding, the vacuum state $|0\rangle$ is defined by the kernel condition $\hat{a}_{n}|0\rangle=0$. Finally, the full field operator becomes
\begin{equation}
\hat{\Psi}(t,x)=e^{-i\mu t+iv x}[\sqrt{\rho}+\hat{\psi}(t,x)],\label{psioperator}
\end{equation}
where $\hat{\psi}$ is the first component of $\hat{\Phi}$. Denoting by $\Phi_{n}=(f_{n}, h_{n})^{\rm t}$, we have from Eq.~\eqref{qexpansion}
\begin{equation}
\hat{\psi}(t,x)=\sum_{n=1}^{\infty}\left[\hat{a}_{n}f_{n}(t,x)+\hat{a}^{\dagger}_{n}h_{n}^*(t,x)\right].\label{qfield}
\end{equation}  

\subsection{Field modes in the presence of a black hole} 
\label{modes}   

Because the system is stationary {\it at the classical level}, solutions to the field equation can be found in the form $\Phi(t,x)=\exp(-i\omega t)\Phi_{\omega}(x)$, and if $\Phi_{\omega}$ is a solution associated to $\omega$, then $\sigma_1\Phi_{\omega}^*$ is also a solution, associated to $-\omega^*$. In this way, we exhaust all real frequencies in the system spectrum by focusing on $\omega>0$ only. Moreover, we say that the system is unstable if there exists a solution with $\mbox{Im}\ \omega>0$. For each such solution, the spectrum necessarily contains also the frequency $\omega^*$, as guaranteed by the hermiticity of the Hamiltonian \cite{Leonhardt2003}. Thus we need to solve
%
%
\begin{equation}
\omega\sigma_3\Phi_{\omega}=\left(-\frac{\partial_{x}^2}{2}-i\mathfrak{m}_{\rm u}\sigma_3\partial_{x}+\frac{g}{g_{\rm u}}\sigma_4\right)\Phi_{\omega},\label{bogo3}
\end{equation}
for $\mbox{Re}\ \omega, \mbox{Im}\ \omega\geq0$, and $\Phi_{\omega}$ is subjected to Eq.~\eqref{bc1}. Furthermore, wave mechanics techniques \cite{Vincenzo} 
applied to Eq.~\eqref{bogo3} imply that $\Phi_{\omega}$ and its first derivative are also continuous at $x=0$. These two conditions, plus the other two in Eq.~\eqref{bc1} give a total of 8 constraints each field mode must satisfy.

\begin{figure}[h!]
\includegraphics[scale=0.56]{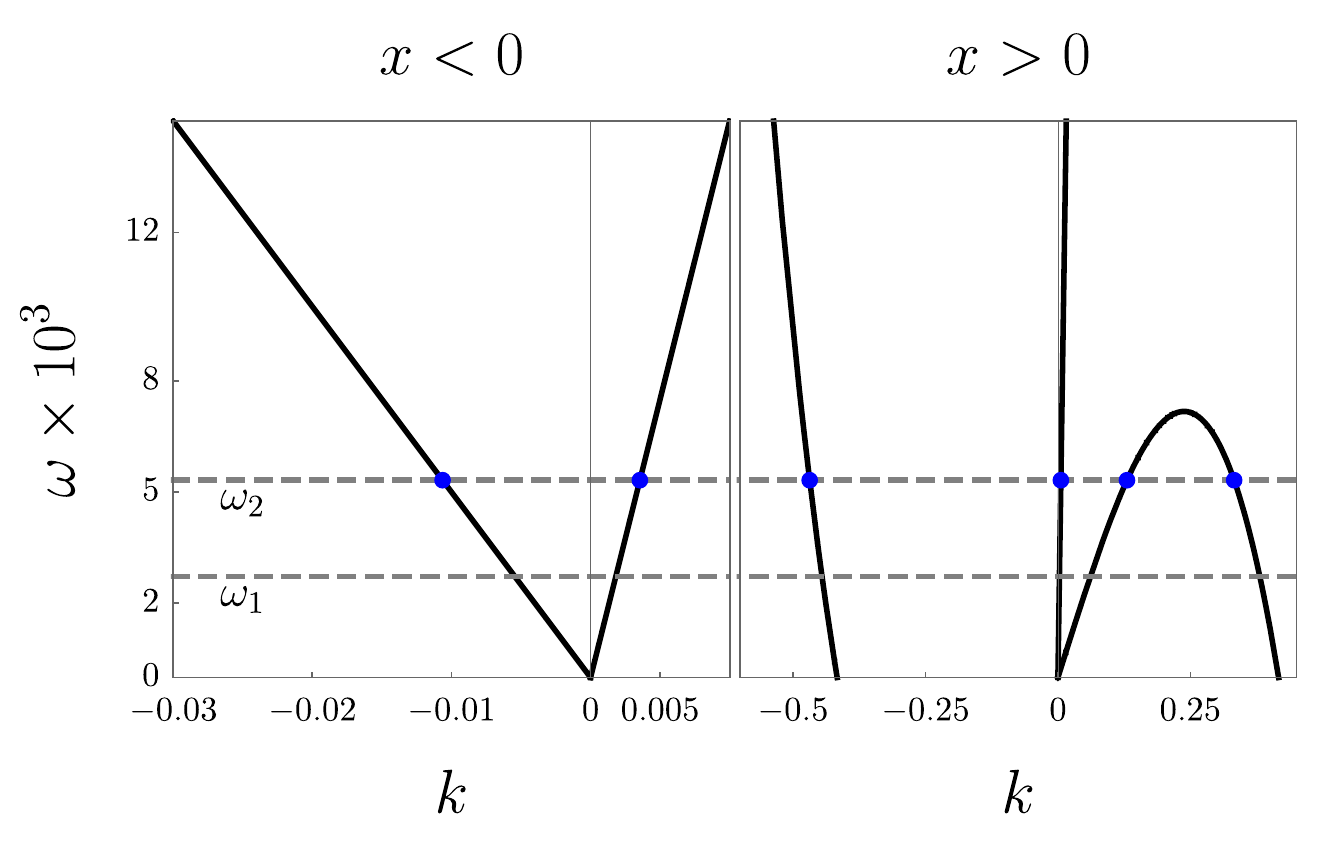}
\caption{Bogoliubov dispersion relation $\omega=\omega(k)$. We set $\mathfrak{m}_{\rm u}=0.5$, $\mathfrak{m}_{\rm d}=1.1$, or $g_{\rm d}/g_{\rm u}\sim 0.2$. Left: dispersion relation in the region $x<0$. The gray dashed lines correspond to the eigenfrequencies within the plot range for a condensate with $-\ell_{1}/2=\ell_{2}/2=60$. The blue points indicate the real solutions for $k$ for $\omega_{2}$.  Right: dispersion relation for the region $x>0$. Because $\mathfrak{m}_{\rm d}>1$, the negative branch of the dispersion relation presents a local maximum. We note that the field modes $\omega_{1}$ and $\omega_{2}$ are below this local maximum.}
\label{figmain2}
\end{figure}

For $x\neq0,-\ell_1/2$ and $\ell_2/2$, the general solution of the ordinary differential equation \eqref{bogo3} is a combination of exponentials of the form $\exp(ikx)\zeta_k$, for constant $\zeta_k$, which, upon substitution in Eq.~\eqref{bogo3} results in the familiar Bogoliubov dispersion relation
\begin{equation}
\left(\omega-\mathfrak{m}_{\rm u}k\right)^2=k^2\left(\frac{g}{g_{\rm u}}+\frac{k^2}{4}\right).\label{disp}
\end{equation}
Notice that for each value of $\omega$, this equation, being a fourth order polynomial equation in $k$, always has 4 solutions (not necessarily distinct), as shown diagrammatically in Fig.~\ref{figmain2}. For simplicity, we shall denote by $p$ the downstream solutions, and thus the general solution for $\Phi_{\omega}$ has the form
\begin{align}
\Phi_{\omega}=\left\{
\begin{array}{c}
\sum_{p}s_{p}e^{ipx}\zeta_{p},\ x>0,\\
\sum_{k}s_{k}e^{ikx}\zeta_{k},\ x<0,
\end{array}\right.\label{gensol}
\end{align}
where the various coefficients $s_{k}$ and $s_{p}$ are integration constants, and
\begin{align}
\zeta_{k}=&\left(\begin{array}{c}
g/g_{\rm u}\\
\omega-\mathfrak{m}_{\rm u}k-k^2/2-g/g_{\rm u}
\end{array}\right).\label{normchannel}
\end{align}

For each possible field mode, clearly at least one of the coefficients $s_{k}$, $s_{p}$ is nonzero, which can then be taken as a normalization constant. This means that a total of 7 conditions are necessary and sufficient to fix the integration constants, which can be done in a straightforward manner by using 7 of the 8 boundary conditions. The remaining equation thus becomes an analytical function of $\omega$ through the roots $k$, $p$, $s_k$'s and $s_p$'s, and whose zeros determine which $\omega$ are in the system spectrum. Furthermore, because the remaining equation dependence on $\omega$ is analytic, the spectrum is discrete. Therefore, the recipe just presented exhausts all possible field modes, which can then be normalized and added to the field expansion of Eq.~\eqref{qexpansion}.

\section{Black hole lifetimes}
\label{bhl}

As in any experimental realization of a trapped BEC, the system confinement implies that small disturbances propagating over the BEC also stay trapped throughout the system evolution. In our analogue model, this property is captured by the Dirichlet boundary conditions \eqref{bc1} at the system (hard) walls, which are a particular way of modeling perfect mirrors for the system radiation. Therefore, considerable differences 
as regards the predictions of finite size and infinitely extended analogue black hole models are expected to occur, for during the black hole existence the energy (continuously) extracted from the background by the Hawking-like process is not allowed to radiate away in the finite size model and stays contained within the hard-walled box. This characteristic, however, does not necessarily lead to {\it dynamical instabilities}, i.e., complex frequencies in the system spectrum. In fact, analogue models containing black hole/white hole pairs usually present a black hole lasing effect \cite{Finazzi}, but in \cite{Garay2001} the authors also found dynamically stable configurations in black hole/white hole analogues in toroidal (and thus finite size) condensates. 

It is noteworthy that because of the complexity of black hole/white hole analogues, the mechanisms leading to stabilization cannot in general be easily disentangled from one another \cite{Jain2007,Tettamanti2016}. In our single black hole analogue model a similar interplay between the system finite size and the radiation process is observed, which leads to dynamically stable or unstable configurations 
rather sensitively depending on the system parameters. For instance, it is an easy exercise to show that if in the configuration of Fig.~\ref{figmain1} if we let $\mathfrak{m}_{\rm d}=\mathfrak{m}_{\rm u}$, although it is stable for finite $\ell_1$, we have instability when $\ell_1\rightarrow\infty$. This analysis shows that subtle finite size effects play a prominent role in the (de-)stabilization of analogue black holes from BECs.

In dynamically unstable scenarios, a natural notion is therefore that of black hole lifetime. In principle, the system spectrum, which contains the complex frequencies, depends solely on the condensate configuration, and thus the involved instability time scales are uniquely determined {\it as long as the condensate exists in that state}, i.e., as long as quantum fluctuations remain small. Therefore, by ensuring that we always start from a scenario of 
well-defined quantum fluctuations (e.g., by determining the condensate depletion), the unstable frequencies set the time scales for the black hole existence which we study in this section.

Four parameters are necessary to specify the analogue black hole: $\{\mathfrak{m}_{\rm u},\mathfrak{m}_{\rm d},\ell_1,\ell_2\}$. For the sake of simplicity, let us assume that one fixes $\ell_1=\ell_2=\ell$. 
%
%
%
\begin{figure}[t]
\includegraphics[scale=0.42]{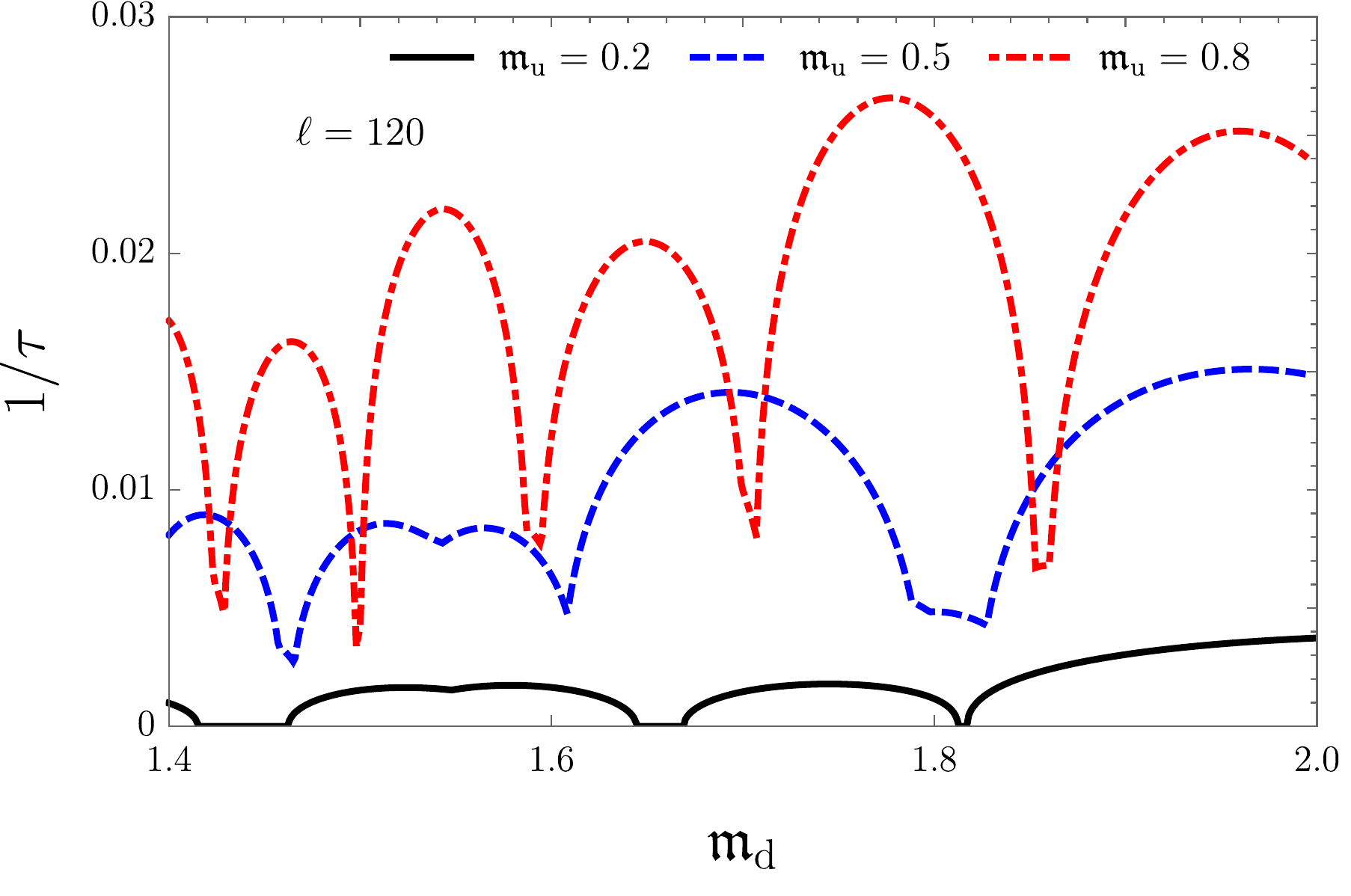}
\caption{Black hole lifetimes as function of the downstream Mach number $\mathfrak{m}_{\rm d}$ for several choices of $\mathfrak{m}_{\rm u}$. Here we set $\ell=120$. Notice that generally no monotonic behavior is observed. Moreover, the lifetimes diverge ($1/\tau=0$) for $\mathfrak{m}_{\rm u}=0.2$ and some values of $\mathfrak{m}_{\rm d}$, i.e., the black continuous curve then touches the $\mathfrak{m}_{\rm d}$ axis.} 
\label{figmain9}
\end{figure}  

We define the {\it analogue black hole lifetime} generally as 
\begin{equation}
\tau=1/(2\mbox{Im}\ \Omega_{\rm max}), 
\label{deftau}
\end{equation}
where $\Omega_{\rm max}$ is the complex frequency in the spectrum with the largest imaginary part. We can study the lifetimes as function of the three free parameters. For instance, in Fig.~\ref{figmain9} we plot $1/\tau$ as functions of the downstream Mach number for several upstream Mach numbers. From the figure, the remarkable effect of the finite system size on the Hawking process can be clearly seen. 
For a system with arbitrarily large size, no dynamical instability exists, and the black hole lifetime is infinite, although the condensate is destroyed by phase fluctuations and is completely depleted \cite{Hohenberg}. On the other hand, when the system has finite size no monotonic behavior as a function of the Mach numbers is observed. Even more noteworthy is the existence of stability regions, which from Fig.~\ref{figmain9} appear for the parameters $\mathfrak{m}_{\rm u}=0.2$, $\ell=120$ (the black continuous curve) where the lifetime diverges. For these parameters, as discussed in the above, there exists a compensation between different mechanisms in the system, that therefore becomes dynamically stable.   
\begin{figure}[t]
\includegraphics[scale=0.42]{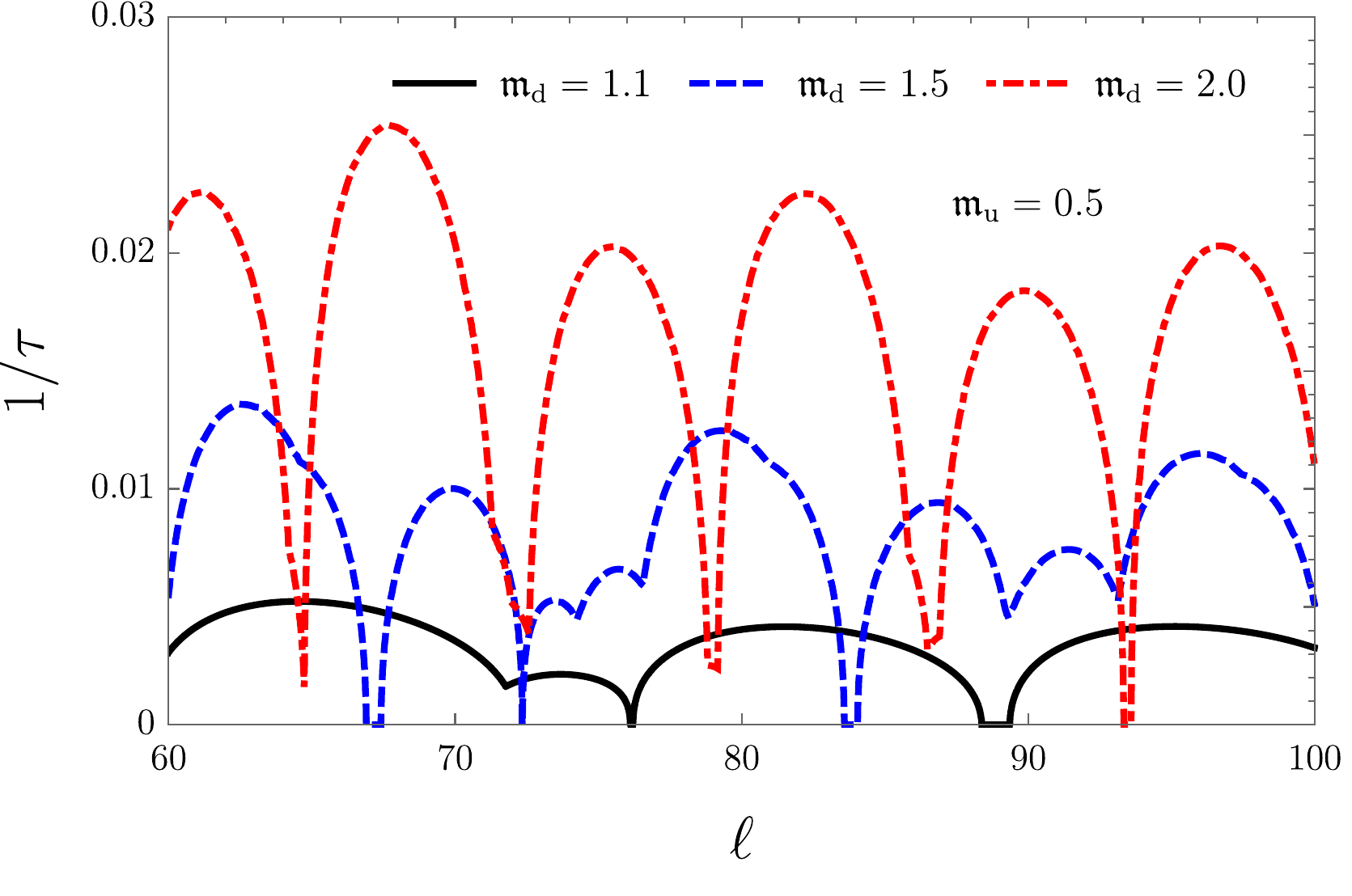}
\caption{Black hole lifetimes as function of the system size for the fixed upstream Mach number $\mathfrak{m}_{\rm u}=0.5$ and several choices of $\mathfrak{m}_{\rm d}$. Similarly to what is observed in Fig.~\ref{figmain9}, there is no clear functional dependence of the lifetimes on the system size. Also, stability regions in the 
space of parameters exist.} 
\label{figmain10}
\end{figure}  
We also depict in Fig.~\ref{figmain10} the lifetimes as a function of the system size. Again, no clear functional dependence with the system parameters can be inferred. We note, however, that the complex dependence of $\tau$ with $\{\mathfrak{m}_{\rm u},\mathfrak{m}_{\rm d}, \ell\}$ is expected because each field mode is built by combining the 8 distinct channels solutions of Eq.~\eqref{disp} and the 8 boundary conditions discussed in subsection \ref{modes}.

\section{The system vacuum state in the presence of instabilities}
\label{vacuum}

Field quantization in the presence of instabilities is a well studied topic \cite{Leonhardt2003,Coutan2010,Ribeiro2020,Lima2010,Lima20102}, and the canonical procedure of section \ref{cq} also then works in general. For our particular goal of simulating quantum depletion, a major aspect of this particular brand of quantization is the notion of an instantaneous vacuum state, a state such that $\langle\hat{\psi}\rangle=0$. As pointed out in \cite{Coutan2010}, the presence of instabilities during the black hole existence prevents the selection of a preferred instantaneous vacuum state, which in stationary configurations, as the name suggests, can be chosen with respect to the laboratory frame via a complete set of normalizable solutions whose positive norm field modes $\Phi_{n}$ are stationary (eigenfunctions) with respect to the generator of time translations $i\partial_t$. In this work we refer to this vacuum state as the quasiparticle vacuum. Notwithstanding, in the presence of instabilities, the operator $i\partial_t$ has non-normalizable eigenfunctions with complex eigenvalues. Thus, bona fide normalizable positive norm field modes constructed from non-normalizable eigenfunctions cannot be eigenfunctions of $i\partial_t$, and in this case we say that quantization spontaneously breaks the time translation symmetry of the theory, and hence no quasiparticle (preferred) vacuum exists. We quote \cite{Macher} for further details regarding the quasiparticle vacuum in infinitely extended 1D quasicondensate analogues. 

In order to highlight the issues with fixing and instantaneous vacuum during an unstable black hole evolution, 
let $\Omega,\ \Omega^*$ be one of the complex frequency pairs in the spectrum, and let the corresponding solutions to Eq.~\eqref{bogo} be $\exp(-i\Omega t)\Phi_{\Omega}(x)$ and $\exp(-i\Omega^* t)\Phi_{\Omega^*}(x)$, respectively. Thus, it follows from the time independence of Eq.~\eqref{scalar} and $\mbox{Im}\ \Omega>0$ that these two solutions have zero norm, but $\langle\Phi_{\Omega},\Phi_{\Omega^*}\rangle:=\lambda\exp(i\theta)\neq0$, in such a way that the two combinations
\begin{align}
&\Phi^{(+)}_{\Omega,\alpha\beta}=\frac{\alpha}{\sqrt{\lambda}}\left[e^{-i\Omega t}\Phi_{\Omega}+\left(\frac{1}{2\alpha^2}+i\beta\right)e^{-i\theta-i\Omega^* t}\Phi_{\Omega^*}\right],\label{umode1}\\
&\Phi^{(-)}_{\Omega,\alpha\beta}=\frac{\alpha}{\sqrt{\lambda}}\left[e^{-i\Omega t}\Phi_{\Omega}-\left(\frac{1}{2\alpha^2}-i\beta\right)e^{-i\theta-i\Omega^* t}\Phi_{\Omega^*}\right],\label{umode2}
\end{align}    
for $\alpha>0$ and real $\beta$ are orthonormal, with $\Phi^{(+)}_{\Omega,\alpha\beta}$ (respectively~$\Phi^{(-)}_{\Omega,\alpha\beta}$) being a positive (respectively~negative) norm solution. If $\Omega$ lies on the imaginary axis, we can add $\Phi^{(+)}_{\Omega,\alpha\beta}$ and $\sigma_1\Phi^{(+)*}_{\Omega,\alpha\beta}$ as a positive-negative norm pair of field modes to the field expansion, whereas if $\Omega$ is not on the imaginary axis, we must add $\Phi^{(+)}_{\Omega,\alpha\beta}$ and $\sigma_1\Phi^{(-)*}_{\Omega,\alpha\beta}$ and the corresponding negative norm counterparts to the expansion. Now, inspection of the modes \eqref{umode1} and \eqref{umode2} reveals that each choice of $(\alpha,\beta)$ is equally acceptable, and it gives rise to a distinct quantum field theory as can be seen by determining the Bogoliubov transformation between the different sets of modes. In particular, we note that it is in principle possible that the vacuum state under study represents a strongly depleted condensate, rendering the whole Bogoliubov expansion inconsistent. We shall return to this question in the next section when we discuss condensate depletion.

\subsection{Quenching to a black hole}
\label{quench}

An elegant way of fixing a preferred vacuum state if the analogue is dynamically unstable is provided by the fact that amongst the various condensate configurations included in our analysis, there are stationary configurations which can be used as initial conditions before quenching  
to the final black hole configuration. For instance, we can start from a system for which, at $t<0$, both Mach numbers $\mathfrak{m}_{\rm u}$, $\mathfrak{m}_{\rm d}$ are smaller than 1, and at $t=0$, the coupling $g_{\rm d}$ is adjusted to set $\mathfrak{m}_{\rm d}>1$ to the required value for created the sonic horizon. 
In this way, as we work in the Heisenberg picture, the initial stationary vacuum remains well defined throughout the system evolution. Specifically, the quantum field $\hat{\Phi}$ has the general expansion of Eq.~\eqref{qexpansion}, 
%
%
where all functions $\Phi_{n}(t,x)$ have positive norm and are solutions of the BdG equation at all times, such that for $t<0$,
\begin{equation}
\Phi_{n}(t,x)=e^{-i\nu_{n}t}\Phi_{\nu_{n}}(x),
\end{equation}
$\nu_{n}>0$ for all $n$. Accordingly, because of the quench, the field mode $\Phi_{n}(t,x)$ after $t=0$ can be expanded in terms of any complete set of solutions, i.e.,
\begin{align}
\Phi_{n}=&\sum_{m=1}^{\infty}\left[\alpha_{n,m}e^{-i\omega_m t}\Phi_{\omega_m}+\beta_{n,m}e^{i\omega_m t}\sigma_1\Phi^{*}_{\omega_m}\right]\nonumber\\
&+\sum_{j}\gamma_{n,j}e^{-i\Omega_j t}\Phi_{\Omega_j},
\end{align}
and the sum in $j$ runs over all complex frequency solutions. Returning to the BdG equation, we conclude from the term $i\partial_t\Phi_{n}$ that $\Phi_{n}(t,x)$ is continuous at $t=0$, which amounts to 
the Fourier expansion
\begin{equation}
\sum_{m=1}^{\infty}\left[\alpha_{n,m}\Phi_{\omega_m}+\beta_{n,m}\sigma_1\Phi^{*}_{\omega_m}\right]+\sum_{j}\gamma_{n,j}\Phi_{\Omega_j}=\Phi_{\nu_{n}}.
\end{equation}
Thus by projecting this equation onto the direction of the field modes the matrices $\alpha_{n,m}$, $\beta_{n,m}$, and $\gamma_{n,j}$ are uniquely fixed, i.e., the solution to the BdG equation is fixed. By using this quantum field expansion instead of the instantaneous quantization when already residing within the unstable phase, the vacuum state is defined to be the quasiparticle vacuum $\hat{a}_{n}|0\rangle=0$, which has a clear interpretation as it is uniquely defined. 

\section{Quantum depletion and the validity of the Bogoliubov expansion}

With the aid of the quantum field expansion, we are able to compute quantum depletion, defined as the vacuum expectation value $\delta\rho=\langle\hat{\psi}^\dagger\hat{\psi}\rangle$. 
The interpretation of quantum depletion is that even at $T=0$, a fraction of the condensed particles, specified by $\delta\rho$, leaves the condensate due to the inherent quantum fluctuations caused by 
the interaction of the particles constituting the system \cite{pita2003bose}. Thus, this {\em measurable} quantity represents a fundamental tool in the theory of Bose-Einstein condensation. Its knowledge is necessary to validate the Bogoliubov expansion, as the ratio depleted/condensed particles should be small for the very expansion implemented in [Eq.~\eqref{psioperator}] to be 
consistent. 

The Bogoliubov expansion applicability criterion, namely, depletion must remain small, does not rigorously fix upper bounds for depletion, and different criteria can be adopted depending on each particular system. For instance, the simulations that follow are such that the largest number of depleted particles occurs near the analogue event horizon, roughly when (reinstating units for clarity) $\xi_{\rm u}\delta\rho\sim2$. For a condensate which has $\xi_{\rm u}\rho\sim60$, this corresponds to $3\%$ of depleted particles near the event horizon. In the present work we fix by convention that the Bogoliubov theory predictions are considered to be accurate as long as depletion remains below $10\%$.

Based on the discussion of section \ref{cq}, we use Eq.~\eqref{qfield} to write the depletion as
\begin{equation}
\delta\rho(t,x)=\sum_{n=1}^{\infty}|h_{n}(t,x)|^2.\label{dep}
\end{equation}

\subsection{Depletion before the black hole formation}

\begin{figure}[b]
\includegraphics[scale=0.42]{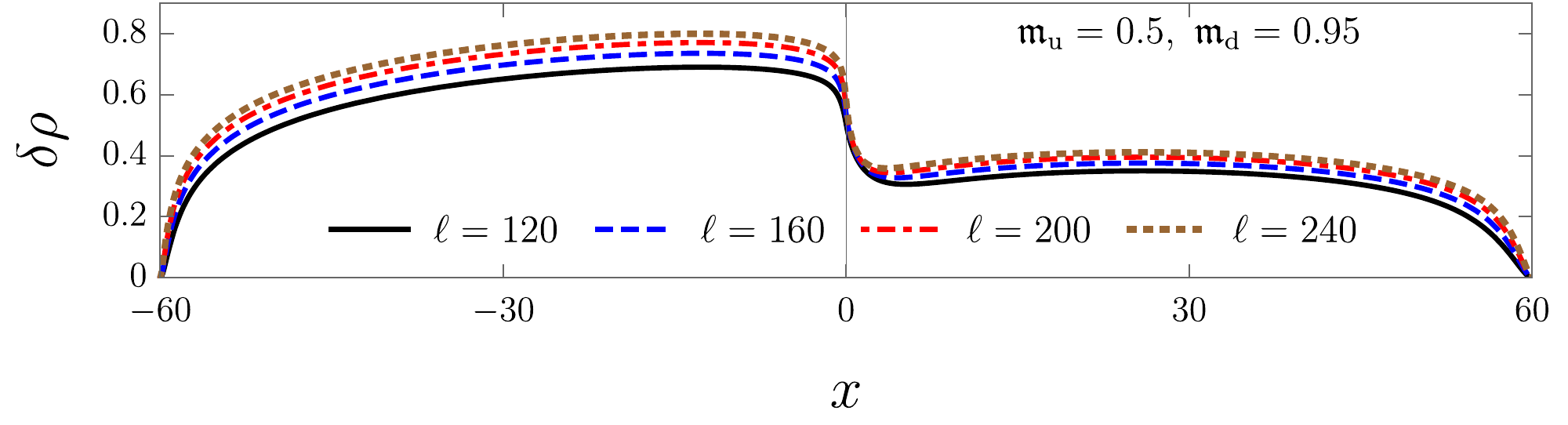}
\caption{Quantum depletion for several condensate sizes in the absence of a black hole $\mathfrak{m}_{\rm d}=0.95$. Here, $\ell_{1}=\ell_{2}$, and we recall that $\ell\equiv(\ell_2+\ell_1)/2$. The curves are scaled in $x$ to fit in the same plot. The effect of the system size is to increase the overall depletion logarithmically.} 
\label{figmain3}
\end{figure}

From Eq.~\eqref{dep}, depletion can be calculated in a straightforward manner using the field modes constructed in subsection \ref{modes}. We recall that a logarithmic divergence is expected to occur in this model as the system size is taken to infinity (see the quantization of \cite{Pavloff2012}, for instance). We thus expect to see an overall increase in depletion as the system size grows, and this is verified in our model already 
{\it when the black hole does not exist}, as shown in Fig.~\ref{figmain3}.   
Other notable features revealed by the plots in Fig.~\ref{figmain3} include their shape robustness as the system size grows, the smaller number of depleted particles at the downstream region, caused by the fact that $g_{\rm d}<g_{\rm u}$ (weaker particle interactions), and the decrease in the number of depleted particles near the condensate boundaries. This latter aspect comes from the particular form of the chosen external potential, which is set such as to impose Dirichlet boundary conditions. From the plot, we can assess that this form of potential results in a vanishing depletion at the condensate walls, a behavior not expected if, for instance, Neumann conditions (no flux into the walls) were adopted.

\subsection{Depletion of a stationary black hole}
\label{stable}

\begin{figure}[b]
\includegraphics[scale=0.4]{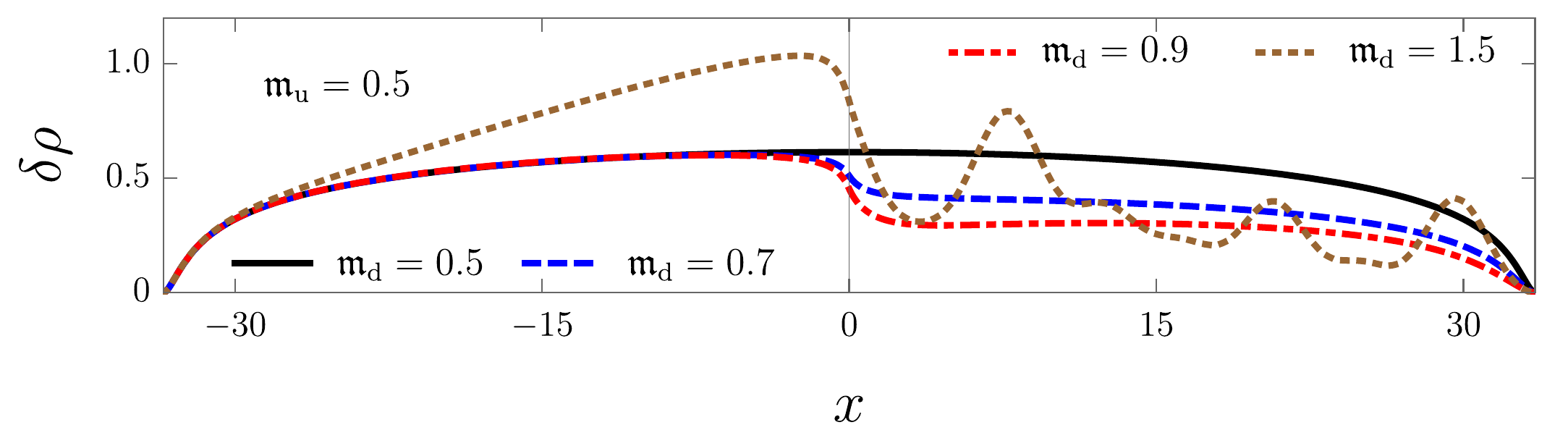}
\caption{Several depletion profiles for fixed $\mathfrak{m}_{\rm u}=0.5$, and $\ell=67$. The continuous black, dashed blue, and dot-dashed red curves correspond, respectively, to $\mathfrak{m}_{\rm d}=0.5$, $\mathfrak{m}_{\rm d}=0.7$, and $\mathfrak{m}_{\rm d}=0.9$, whereas the dotted brown curve depicts depletion for the stationary, stable configuration with $\mathfrak{m}_{\rm d}=1.5$. Deep into the upstream region we see that the sound barrier at $x=0$ leaves no imprint in the black hole's absence, but as the analogue event horizon forms, the upstream noncondensed cloud changes due to the analogue black hole Hawking radiation. We also note the intricate depletion behavior at the downstream region after the black hole formation, which is 
in sharp distinction to the featureless depletion profile without black hole.} 
\label{figmain11}
\end{figure}

When the black hole is formed, the depletion curves are qualitatively distinct, and let us consider first quantum depletion in stationary (dynamically stable) black hole configurations, as discussed in section \ref{bhl}. We present in Fig.~\ref{figmain11} our findings for a stationary black hole analogue defined by $\mathfrak{m}_{\rm u}=0.5$, $\mathfrak{m}_{\rm d}=1.5$, $\ell=67$, which, from Fig.~\ref{figmain10}, can indeed be seen to correspond to a divergent lifetime. 
Figure \ref{figmain11}, which represents one of our major findings, depicts how the upstream noncondensed cloud outside the black hole $(x<0)$ is affected by the Hawking-like radiation. We also plot depletion profiles for a fixed upstream Mach number $\mathfrak{m}_{\rm u}=0.5$ and different $\mathfrak{m}_{\rm d}<1$. The latter reinforces the intuition in the absence of a black hole, a variable $g$ at $x=0$ models a sound barrier for the phonon field that should not be perceived far away from the barrier ($x\ll1$). The continuous, dashed, and dot-dashed curves in Fig.~\ref{figmain11} show that in the black hole's absence, depletion is indeed only locally 
affected by the sound barrier, and for if $x\ll1$, it is not possible to detect its presence by measurements of the depleted cloud. However, after the black hole is formed, a clear contribution to depletion deep into the upstream region appears.

 It is possible to directly correlate the imprint on the upstream noncondensed cloud asymptotically far from the analogue event horizon with to the Hawking-like radiation if we assume that the condensate is extremely elongated, by using, for instance, the field modes of \cite{Pavloff2012}. Such a  calculation, however, requires the use of frequency cutoffs to render depletion finite in our quasi-1D setup; such cutoffs can be inferred from our finite size model. We notice also from Fig.~\ref{figmain11} the intricate equilibrium pattern displayed by the depleted cloud in the downstream region. We discuss how to probe the event horizon existence by measurements of the downstream depleted cloud in the next subsection.

\subsection{Depletion after the formation of an unstable black hole}

Now we consider the most common case of analogues that can be studied with our confined system: 
Dynamically unstable black holes. When an analogue black hole has just formed, and a phonon field instability then develops, we expect to see a continuous extraction of atoms from the condensate (depletion increase), and the whole system will eventually assume a new configuration. Naturally, to determine how the system will 
ultimately stabilize and to describe the nature of the final state, a fully self-consistent backreaction analysis is required, which is beyond the scope of this work. Nevertheless, the instability onset can be explored with our quantization scheme, and this subsection is dedicated to it.

As discussed in section \ref{vacuum}, in the absence of a 
stationary regime in unstable scenarios, we need to specify initial conditions for the system, and for the sake of illustration, let us therefore start by considering an instantaneous vacuum state for the phonon field, when the black hole is already formed. Note that even if the unstable modes possess negligible absolute frequencies, it is not in general possible to treat the system as an effectively stable one due to the breakdown of time translation symmetry (see in this regard section \ref{vacuum}).

\begin{figure}[t]
\includegraphics[scale=0.6]{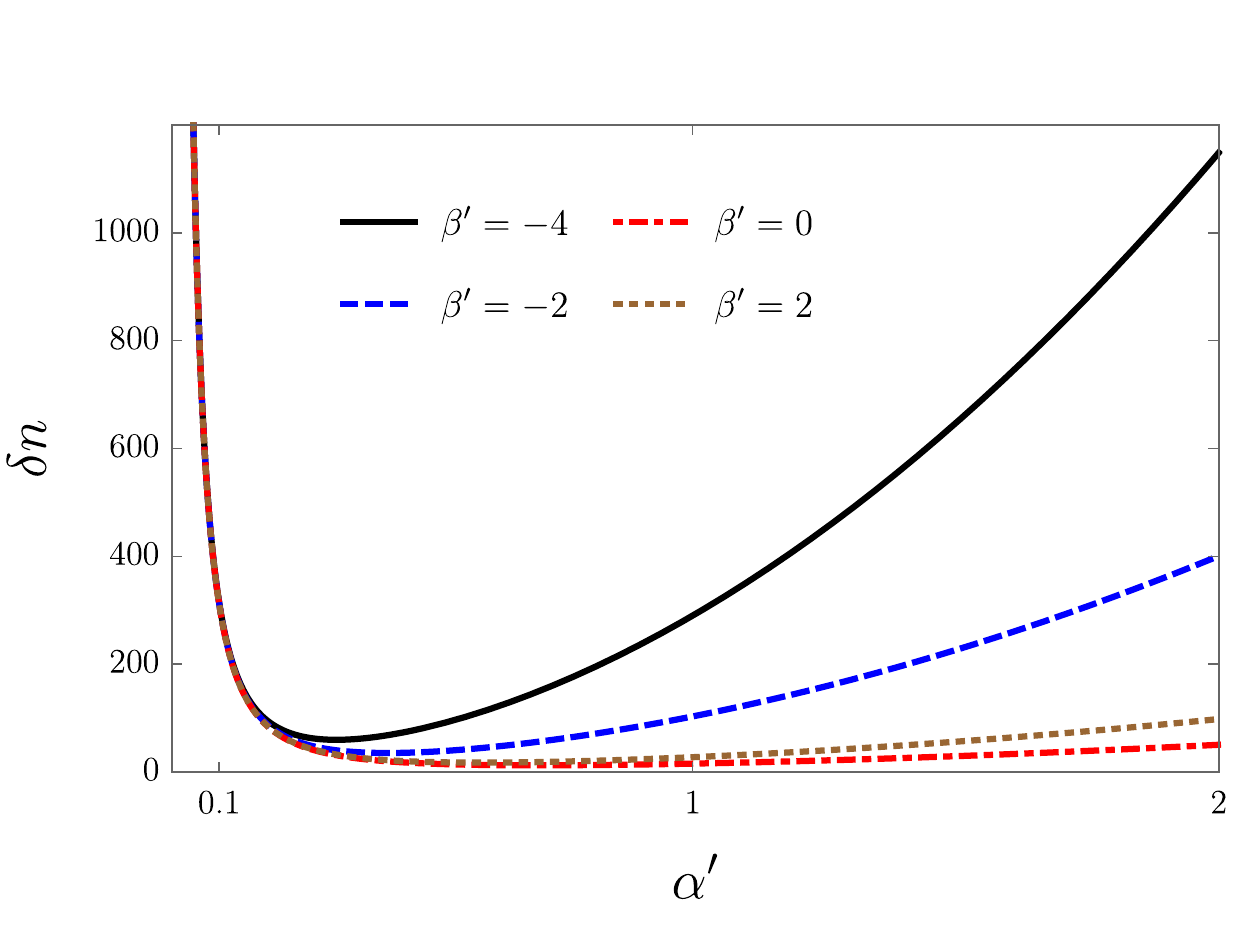}
\caption{Contribution to the quantum depletion coming from different choices of instantaneous vacuum states for the sector $\Omega_2$. System parameters are $\mathfrak{m}_{\rm u}=0.5$, $\mathfrak{m}_{\rm u}=1.1$, and $\ell=120$.} 
\label{figmain5}
\end{figure}

In order to gain further insight,  
we shall treat an explicit example thoroughly. 
Consider the case where $\mathfrak{m}_{\rm u}=0.5$, $\mathfrak{m}_{\rm d}=1.1$, and $\ell_{1}=\ell_{2}=120$, which corresponds to a condensate of total size $\ell=120$. The perturbation spectrum for this configuration contains exactly six complex frequencies, obtained from $\Omega_1\sim i8\times10^{-4}$ and $\Omega_2\sim (70.76 + 3 i)\times10^{-4}$. Therefore, from the discussion that leads to Eqs.~\eqref{umode1} and \eqref{umode2}, we see that the space of possible choices for the system vacuum is parametrized by 4 real parameters, two for each complex frequency in the upper right portion of the complex plane. For each vacuum state, the sum in Eq.~\eqref{dep} splits into a 
time-dependent contribution to the system depletion (which contains the the unstable field modes), and a time-independent part denoted by $\delta\rho_{\rm s}$, the latter also being independent of the vacuum choice. Following the notation of Eq.~\eqref{qfield}, the depletion assumes the form
\begin{equation}
\delta\rho=\delta\rho_{\rm s}+\left|h^{(+)}_{\Omega_1,\alpha\beta}\right|^2+\left|h^{(+)}_{\Omega_2,\alpha'\beta'}\right|^2+\left|f^{(-)}_{\Omega_2,\alpha'\beta'}\right|^2,\label{depexample}
\end{equation}
where $\alpha,\alpha'>0$, $\beta,\beta'$ are {\it any} real parameters. The importance of this result for our analysis is that however large the time scale imposed by the instability for the condensate depletion is, it is in principle possible that no vacuum state exists for which $\delta\rho\ll\rho$, rendering the 
whole quantization procedure based on the Bogoliubov expansion inconsistent. We can visualize this by counting the total number of depleted particles, $\delta N=\int \d x \delta\rho$, which in view of Eq.~\eqref{depexample} splits into a contribution from the stable modes, a contribution from the sector $\Omega_1$, and one from the sector $\Omega_2$. We plot in Fig.~\ref{figmain5} the number of depleted particles due to the unstable modes from the sector $\Omega_2$ for different choices of initial states parametrized by
$\alpha',\beta'$.

Inspection of Fig.~\ref{figmain5} reveals that depending on the parameters $\alpha'$, $\beta'$, 
the predictions of Bogoliubov theory can not be expected to be completely reliable. For instance, for a system with a total of 6000 particles, 1000 depleted particles correspond to 16\% of the particles not in the condensed phase. This violates the small depletion criterion of the Bogoliubov expansion, and it is thus not possible to decide whether the quantization is consistent, or even if the corrections to the condensate remain negligible. A regime of ``initially'' large depletion correspond to cases where the unstable phase already played a relevant role, and de-stabilization processes are taking over the condensate evolution. 

\begin{figure}[b]
\includegraphics[scale=0.42]{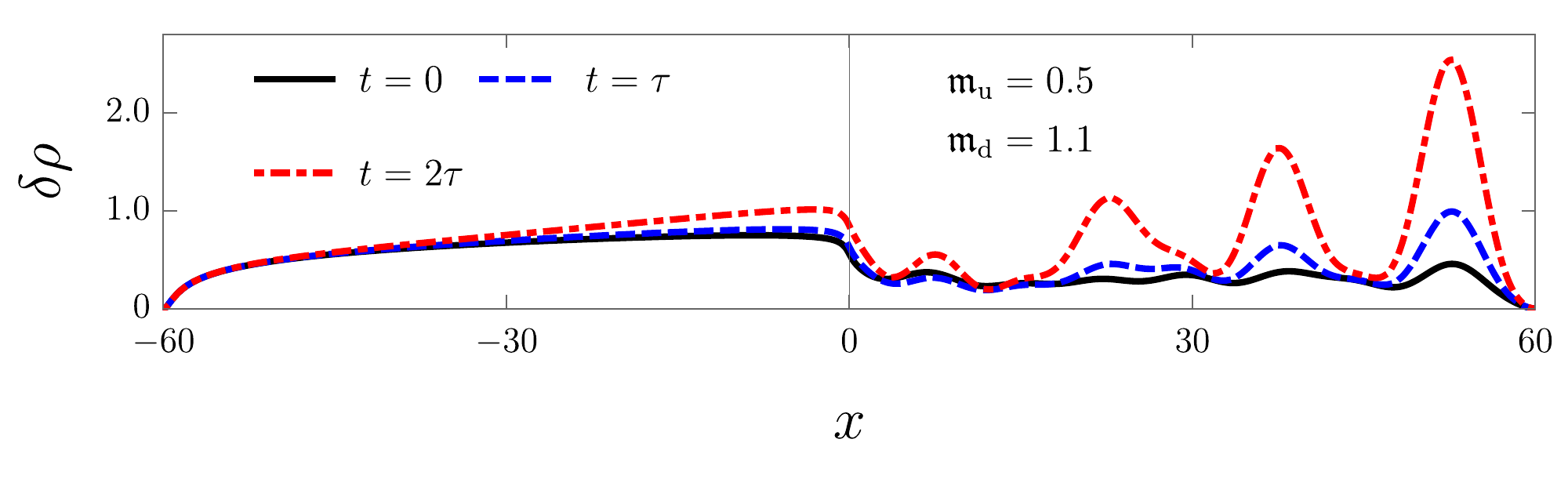}
\caption{Depletion profiles as function of time for a black hole characterized by $\mathfrak{m}_{\rm u}=0.5$, $\mathfrak{m}_{\rm d}=1.1$, and $\ell=120$, with the system in its vacuum state of minimum depletion. Three major features are observed: Initially (black curve) the depletion distribution inside the black hole does not resemble the stable curves of Fig.~\ref{figmain3}; as time passes, the number of depleted particles increases outside the black hole; an oscillatory pattern emerges inside the black hole. Here, the lifetime $\tau$ is defined in \eqref{deftau}.} 
\label{figmain8}
\end{figure}

In such cases, the presently used non-number-conserving Bogoliubov expansion is not suitable, and a full analysis which includes backreaction effects, within a number-conserving analysis, is unavoidable. On the other hand, as the number of depleted particles is bounded from bellow, there must exist a vacuum state which renders  the smallest depletion, as the Fig.~\ref{figmain5} suggests. It is straightforward to show, by minimizing the total number of depleted particles with respect to the parameters $\alpha, \alpha', \beta, \beta'$, that there is only one possible choice for the minimizer, 
which is depicted in Fig.~\ref{figmain8}. 
 This result presents a sharp lower bound for depletion in our black hole analogue.

The problem of identifying a vacuum state during the unstable phase usually restricts the applicability of the theory to the study of asymptotic regimes, and as shown here, in the presence of condensates, the analysis is further complicated by whether the condensate persists against quantum depletion. 
Yet, this problem is of theoretical importance only, as it comes from the assumption that the system is stationary, and everlasting, even lthough spontaneously growing quantum fluctuations break the time translation symmetry. In experimental realizations, the condensate and the black hole setting must have a starting point, which sets the system vacuum (the condensate) throughout its evolution. This is captured by the quench described in subsection \ref{quench}, which we explore now. 
 
\begin{figure}[b]
\includegraphics[scale=0.42]{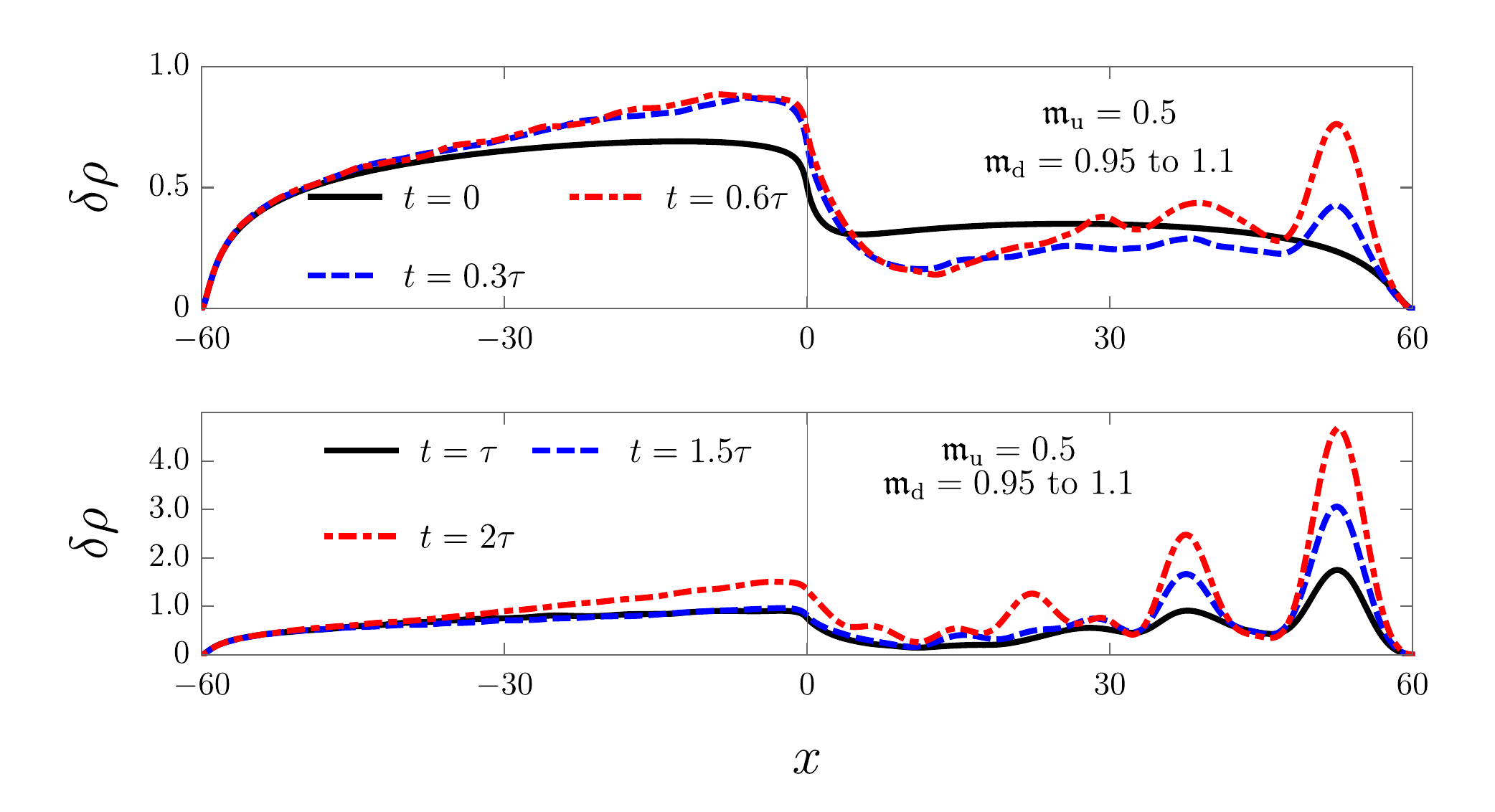}
\caption{Several depletion profiles as function of time for a quenched black hole. The system is set to have $\mathfrak{m}_{\rm u}=0.5$, $\mathfrak{m}_{\rm d}=0.95$, and $\ell=120$ for $t<0$, and we change $\mathfrak{m}_{\rm d}=1.1$ after the quench at $t=0$. } 
\label{figmain4}
\end{figure}

Still assuming the model with $\mathfrak{m}_{\rm u}=0.5$, and $\ell_{1}=\ell_{2}=120$, let us consider the case for which at $t=0$, the system passes from $\mathfrak{m}_{\rm d}=0.95$ (the continuous line in Fig.~\ref{figmain3}) to $\mathfrak{m}_{\rm d}=1.1$. In this case, the lifetime $\tau$ for the system is set by $\Omega_1$, which has the larger imaginary part. 
By using Ref.~\cite{Steinhauer2019} and its 
experimental parameters as a guide, and returning to dimensionful units, we find 
$\tau\approx 8$ s for a chemical potential of $70$ Hz. We plot in Fig.~\ref{figmain4} our findings for this quenched system.

As advocated in the above, by starting from a truly stationary system in its uniquely defined quasiparticle vacuum, we can study the system evolution in a consistent and self-contained way when the Hawking process is switched on. Inspection of Fig.~\ref{figmain4} reveals that as the black hole forms, a nontrivial quantum depletion response is triggered, with the formation of an interference pattern inside the black hole (downstream region) and the continuous increase of the overall number of depleted particles, inside and outside the black hole. For this model, $\mathfrak{m}_{\rm d}$ is increased by decreasing the particle interaction strength $g_{\rm d}$, and thus intuition would suggest that the number of depleted particles for the black hole with $\mathfrak{m}_{\rm d}=1.1$ would be smaller then the one for the plotted one in Fig.~\ref{figmain3} for the  non-black-hole case with $\mathfrak{m}_{\rm d}=0.95$. However, the curves in Fig.~\ref{figmain4} show that the distribution of depleted particles inside the black hole does not follow the ``monotonic'' behavior we see in stable configurations as displayed in Fig.~\ref{figmain3}. We furthermore call attention to the distribution of depleted particles outside the black hole, where we see the emergence of the depletion signal discussed in subsection \ref{stable}.

\begin{figure}[t]
\includegraphics[scale=0.42]{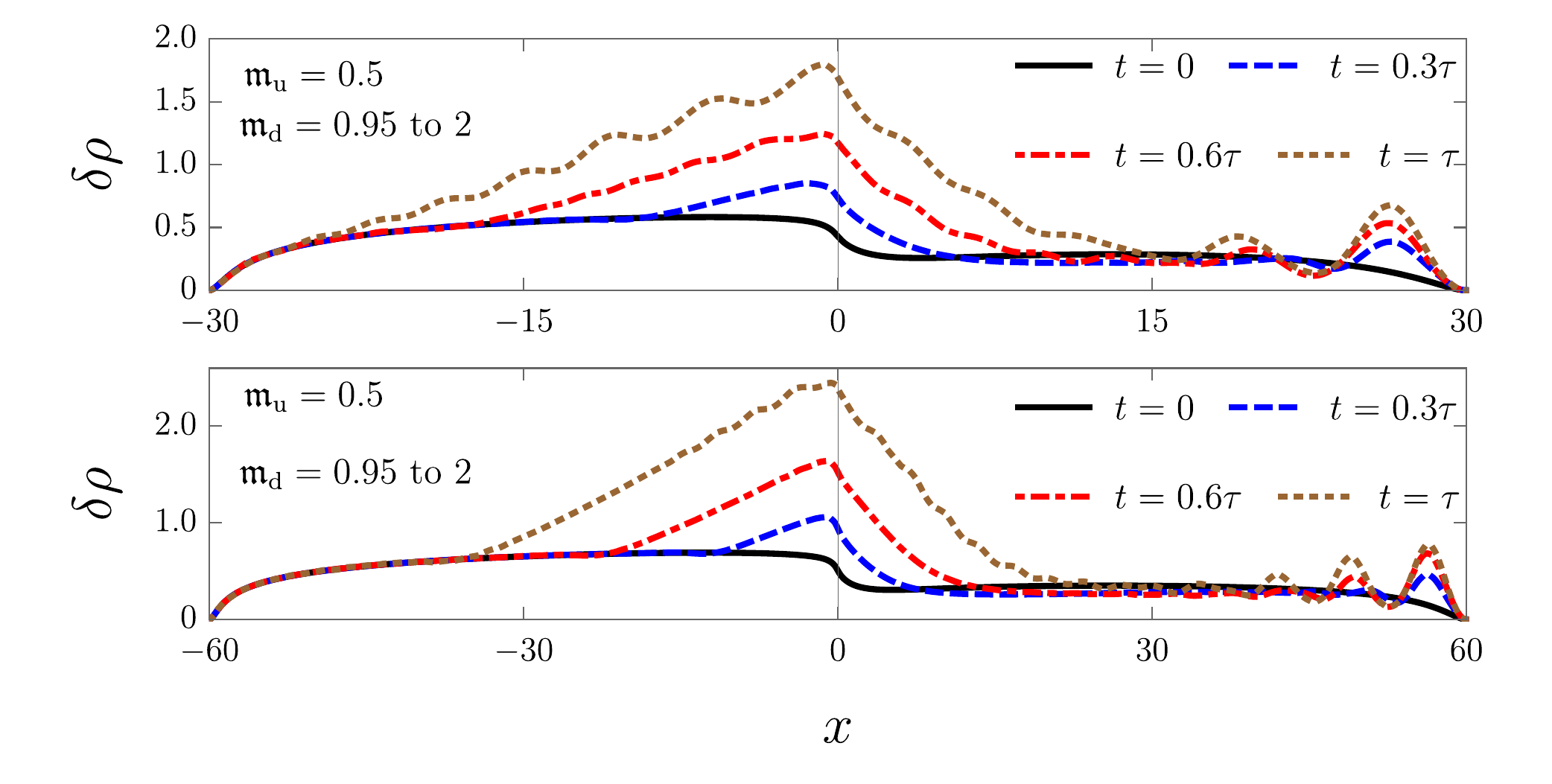}
\caption{Quantum depletion for quenched black holes of different sizes. The black holes have $\mathfrak{m}_{\rm u}=0.5$, and the quench changes $\mathfrak{m}_{\rm d}$ from $0.95$ to $2$ at $t=0$. The increased downstream Mach number leads to a stronger radiation \cite{Pavloff2012}. Upper panel: $\ell=60$. Lower panel: $\ell=120$. Both systems present similar depletion behavior, with the emergence of an oscillatory pattern inside the black hole, and the peculiar upstream-depleted cloud signal discussed in Subsec.~\ref{stable}, which forms at the analogue event horizon ($x=0$) and then propagates against the condensate flow.} 
\label{figmain7}
\end{figure}

We can follow the ramp-up of the Hawking radiation with a better resolution by taking analogue models with higher downstream Mach numbers, which corresponds to stronger radiation  \cite{Pavloff2012}. For the sake of illustration, simulations are depicted in Fig.~\ref{figmain7} for the Mach numbers $\mathfrak{m}_{\rm u}=0.5$, $\mathfrak{m}_{\rm d}=2$.
%
We obtain a lifetime, cf.~Eq.
\eqref{deftau}, of $\tau\sim1$ sec for a system of total size $\ell=120$ and chemical potential $70$ Hz, about ten times smaller than the one found for the system in Fig.~\ref{figmain4} with $\mathfrak{m}_{\rm d}=1.1$. Furthermore, $\tau\sim 0.7$ s for a size $\ell=60$. Figure \ref{figmain7} shows that, as the Hawking-like process is switched on, the cloud of depleted particles increases in a manner directly correlated to the radiated signal: 
The depleted cloud in a fixed upstream region only responds to the radiation as it reaches that region, and as time passes, because the system is out of equilibrium, the number of depleted particles increases gradually until Bogoliubov theory is no longer reliable. Furthermore, although this depletion response to the radiation sheds some light into how the system evolution occurs, it does not uniquely fix how the background condensate changes, i.e., how backreaction takes place. For instance, it is not possible to track down the origin of the upstream depleted particles, whether it was followed by a condensate upstream and/or downstream local depletion. Nevertheless, we reiterate that is it necessary to know the condensate depletion in order to establish any consistent number-conserving analysis.

\subsection{Power spectrum of quantum depletion} 
\begin{figure}[t]
\includegraphics[scale=0.58]{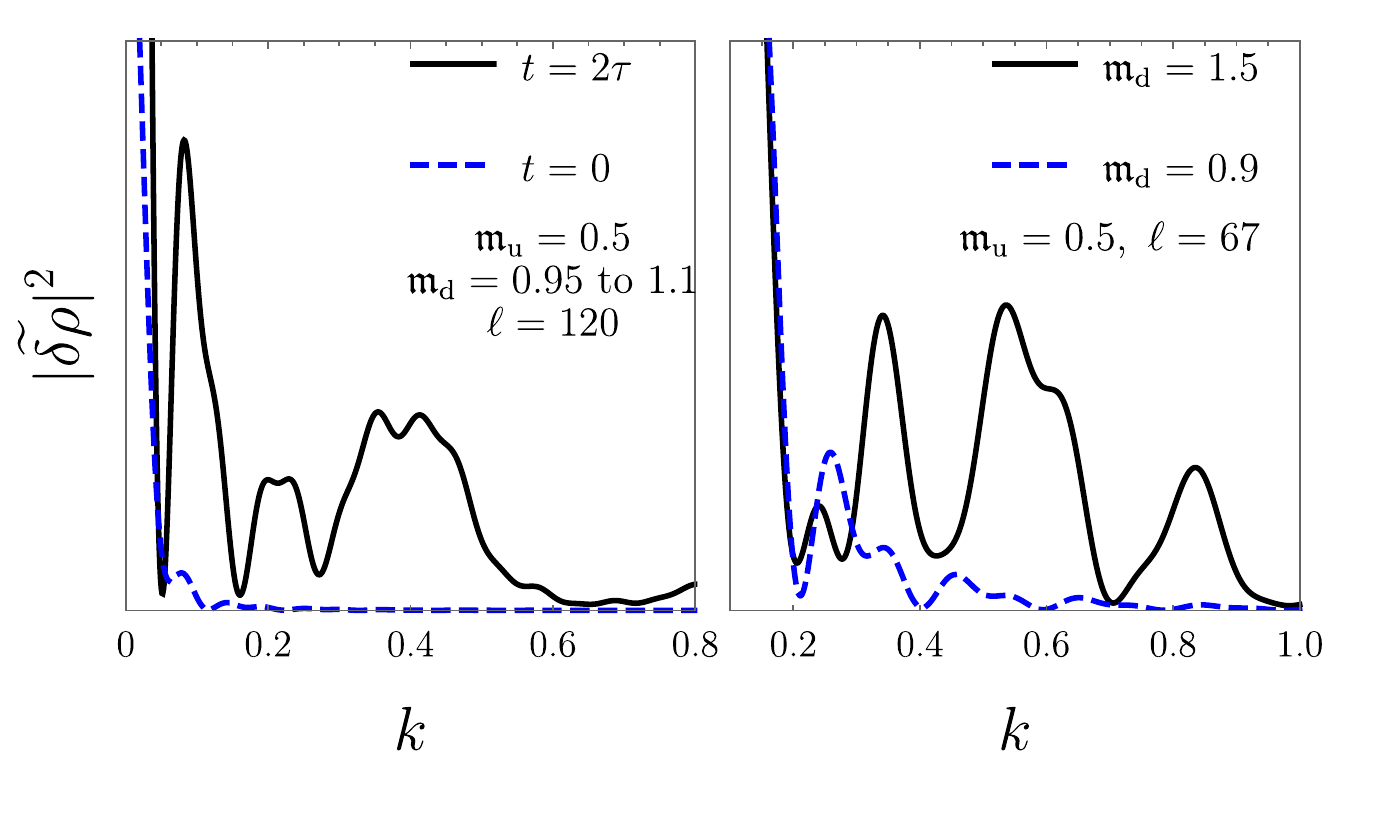}
\caption{Power spectrum of the depletion profile for two black hole analogues. Left panel: the two curves represent the observed spectrum at different instants of time for an unstable black hole configuration, both at the beginning of the quench, and after a time $t=2\tau$, for a black hole characterized by $\mathfrak{m}_{\rm u}=0.5$, $\ell=120$, and $\mathfrak{m}_{\rm d}=0.95$ for $t<0$, $\mathfrak{m}_{\rm d}=1.1$, $t>0$. The black continuous curve shows the formation of a bump near $k\sim0.4$, absent before the black hole forms, as indicated by the blue dashed curve. Right panel: power spectrum for the stable black hole (continuous black curve) of Fig.~\ref{figmain11}. The blue dashed curve shows the power spectrum before the black hole formation.} 
\label{figmain6}
\end{figure}

As a particularly noteworthy feature, our analysis reveals the possibility of probing the existence of an analogue event horizon from the emergent interference pattern manifest in the local quantum depletion
 using the Bragg technique employed by \cite{Hadzibabic2017}. Denoting Fourier transforms as $\widetilde{\rho}(k)=\int\d x\exp(-ikx)\rho(x)$, and similarly for $\widetilde{\delta\rho}(k)$, 
Ref.~\cite{Hadzibabic2017} exploits the fact that in some configurations $\widetilde{\rho}(k)$ decays faster for large $k$ in comparison to the polynomial decay of $\widetilde{\delta\rho}(k)$.
One thus obtains a large $k$ window which is sensitive to depletion. The emergence of the interference pattern in Fig.~\ref{figmain4} upon formation of the horizon
transforms to distinct peaks in $\widetilde{\rho}(k)$,  
as shown for two black hole examples in Fig.~\ref{figmain6}. 

\section{Summary and final remarks}

We proposed a finite size quasi-1D analogue black hole model which contains a single event horizon. The model is possibly the simplest one that encapsulates such a single horizon and enforces zero vacuum fluctuations (Dirichlet boundary conditions) at the condensate walls. We note that other boundary conditions can also be used, as for instance Neumann conditions. However, Dirichlet conditions are better suited when we take into account that in experimental realizations the condensate is subject to confining potentials along its symmetry axis. Furthermore, the main advantage of adopting a finite size condensate is that it allows for a controlled usage of Bogoliubov theory by rendering quantum depletion finite and well defined in a quasi-1D system.

We demonstrate the existence of finite-size-induced dynamical instabilities for the majority of the black hole analogues we probe. By reviewing canonical field quantization in the presence of instabilities, we show that if the phonon field is not carefully quantized, the theory can represent a strongly depleted condensate, which renders the Bogoliubov expansion inconsistent. We addressed this problem by employing a quenching from a stationary configuration in its quasiparticle vacuum to the final black hole under study. This procedure enabled us to simulate the evolution of the depletion cloud during the instability onset. We found that two distinct signatures of the Hawking process emerge when the event horizon forms, the first one being the appearance of an oscillatory pattern in the depletion cloud inside the black hole which translates to distinct peaks in its power spectrum. And the second one is what follows: the quench we impose reveals the existence of a link between the radiation emitted by the black hole and the depletion cloud, namely, the local depletion at a region outside the black hole starts to increase as the radiation reaches that region. This represents a novel signature of the Hawking radiation ramp-up that is related to the overall distribution of particles either belonging to the condensed or noncondensed part of the system.

We finally comment on the relevance of our results for real black holes.
In analogue gravity, the 
quantum many-body wave function of the whole system 
can (in principle) be accessed via the observer in the lab. Similarly, we anticipate that while in the currently existing nonunified theory of quantized matter fields propagating in a classical, fixed curved spacetime background,   
the 
depletion oscillations are hidden behind the horizon, a unitary closed system evolution, potentially provided by a future unification of gravity with the matter fields, will effectively provide access to the black hole quantum 
interior and thus also to the depletion oscillations we have investigated.  

\acknowledgments
This work has been supported by the National Research Foundation of Korea under 
Grants No.~2017R1A2A2A05001422 and No.~2020R1A2C2008103.

%
%

\bibliography{1Dbhv7}

%
%

\end{document}